\documentclass[onecolumn]{aastex62}
\usepackage[section]{placeins}
\usepackage{graphicx}
\usepackage{longtable}
\usepackage{hyperref}
\usepackage{url}
\usepackage{natbib}
\hypersetup{colorlinks=true,citecolor=blue,linkcolor=blue,urlcolor=blue}
\usepackage{comment}
\usepackage{sidecap}
\usepackage{etoolbox}

\newcommand{\eauthor}[2]{\yauthor{\href{mailto: #1}{#2}}}

\newcommand{\STScI}{\affiliation{Space Telescope Science Institute, 3700 San Martin Dr, Baltimore, MD 21218, USA}}

\newcommand{\QUB}{\affiliation{Astrophysics Research Centre, School of Mathematics and Physics, Queen's University Belfast, Belfast BT7 1NN, UK}}

\newcommand{\CfA}{\affiliation{Center for Astrophysics \textbar{} Harvard \& Smithsonian, 60 Garden Street, Cambridge, MA 02138-1516, USA}}

\newcommand{\CIERA}{\affiliation{Center for Interdisciplinary Exploration and Research in Astrophysics and Department of Physics and Astronomy, \\Northwestern University, 2145 Sheridan Road, Evanston, IL 60208-3112, USA}}

\newcommand{\Steward}{\affiliation{Steward Observatory, University of Arizona, 933 North Cherry Avenue, Tucson, AZ 85721, USA}}
\newcommand{\PSUa}{\affiliation{Department of Astronomy \& Astrophysics, The Pennsylvania State University, University Park, PA 16802, USA}}
\newcommand{\PSUb}{\affiliation{Institute for Computational \& Data Sciences, The Pennsylvania State University, University Park, PA 16802, USA}}
\newcommand{\PSUc}{\affiliation{Institute for Gravitation and the Cosmos, The Pennsylvania State University, University Park, PA 16802, USA}}

\newcommand{\UCSD}{\affiliation{Department of Astronomy and Mount Laguna Observatory, San Diego State University, San Diego, CA 92182, USA}}
\newcommand{\JHU}{\affiliation{Physics and Astronomy Department, Johns Hopkins University, Baltimore, MD 21218, USA}}
\newcommand{\Duke}{\affiliation{Department of Physics, Duke University Durham, NC 27708, USA}}
\newcommand{\Baylor}{\affiliation{Department of Physics, Baylor University, One Bear Place 97316, Waco, TX 76798-7316, USA}}

\shorttitle{Roman SLSNe}
\shortauthors{Gomez et al.}

\begin{document}


\title{\Large Characterizing Superluminous Supernovae with Roman}

\correspondingauthor{Sebastian Gomez}
\email{sgomez@stsci.edu}

\eauthor{sgomez@stsci.edu}{Sebastian Gomez}
\STScI

\eauthor{kate.alexander@northwestern.edu}{Kate Alexander}
\Steward

\eauthor{eberger@cfa.harvard.edu}{Edo Berger}
\CfA

\eauthor{peter.blanchard@northwestern.edu}{Peter K. Blanchard}
\CIERA

\eauthor{floor.broekgaarden@cfa.harvard.edu}{Floor Broekgaarden}
\CfA

\eauthor{teftekhari@northwestern.edu}{Tarraneh Eftekhari}
\CIERA

\eauthor{ofox@stsci.edu}{Ori Fox}
\STScI

\eauthor{kiranjyot.gill@cfa.harvard.edu}{Kiranjyot Gill}
\CfA

\eauthor{daichi.hiramatsu@cfa.harvard.edu}{Daichi Hiramatsu}
\CfA

\eauthor{bjoshi5@jhu.edu}{Bhavin Joshi}
\JHU

\eauthor{mkarmen1@jhu.edu}{Mitchell Karmen}
\JHU
 
\eauthor{takashi.moriya@nao.ac.jp}{Takashi Moriya}
\affiliation{National Astronomical Observatory of Japan, National Institutes of Natural Sciences, 2-21-1 Osawa, Mitaka, Tokyo 181-8588, Japan}

\eauthor{matt.nicholl@qub.ac.uk}{Matt Nicholl}
\QUB

\eauthor{rquimby@sdsu.edu}{Robert Quimby}
\UCSD

\eauthor{regos@konkoly.hu}{Eniko Regos}
\affiliation{Konkoly Observatory, Konkoly Thege M. 15-17, Budapest, H-1121 Hungary}

\eauthor{arest@stsci.edu}{Armin Rest}
\STScI

\eauthor{Ben\_Rose@baylor.edu}{Benjamin Rose}
\Duke\Baylor

\eauthor{mshahbandeh@stsci.edu}{Melissa Shahbandeh}
\STScI

\eauthor{ashley.villar@gmail.com}{V. Ashley Villar}
\PSUa\PSUb\PSUc

\begin{abstract}

Type-I Superluminous Supernovae (SLSNe) are an exotic class of core-collapse SN (CCSN) that can be up to 100 times brighter and more slowly-evolving than normal CCSNe. SLSNe represent the end-stages of the most massive stripped stars, and are thought to be powered by the spin-down energy of a millisecond magnetar. Studying them and measuring their physical parameters can help us to better understand stellar mass-loss, evolution, and explosions. Moreover, thanks to their high luminosities, SLSNe can be seen up to greater distances, allowing us to explore how stellar physics evolves as a function of redshift. The High Latitude Time Domain Survey (HLTDS) will provide us with an exquisite dataset that will discover 100s of SLSNe. Here, we focus on the question of which sets of filters and cadences will allow us to best characterize the physical parameters of these SLSNe. We simulate a set of SLSNe at redshifts ranging from z = 0.1 to z = 5.0, using six different sets of filters, and cadences ranging from 5 to 100 days. We then fit these simulated light curves to attempt to recover the input parameter values for their ejecta mass, ejecta velocity, magnetic field strength, and magnetar spin period. We find that four filters are sufficient to accurately characterize SLSNe at redshifts below $z = 3$, and that cadences faster than 20 days are required to obtain measurements with an uncertainty below 10\%, although a cadence of 70 days is still acceptable under certain conditions. Finally, we find that the nominal survey strategy will not be able to properly characterize the most distant SLSNe at $z = 5$. We find that the addition of 60-day cadence observations for 4 years to the nominal HLTDS survey can greatly improve the prospect of characterizing these most extreme and distant SNe, with only an 8\% increase to the time commitment of the survey. \\

\noindent
\textbf{Roman Community Survey:} High Latitude Time Domain Survey

\end{abstract}

\keywords{stellar physics and stellar types -- supernovae -- massive stars \newpage}

\section{Superluminous Supernovae}\label{sec:slsn}

Type Ib/c supernovae (SNe) are a common type of core-collapse SN (CCSN) that result from the explosions of massive stars which have lost their hydrogen (SNe~Ib) and helium (SNe~Ic) envelopes. These SNe are known to be powered by the radioactive decay of $^{56}$Ni \citep{Filippenko97}, but the mechanism by which their progenitors lose their envelopes is not well understood, ranging from stellar winds, to interaction with a binary companion, or pair-instability events (e.g., \citealt{Podsiadlowski92, Woosley07, Aguilera-Dena23}). Over the past decade, a rare class of stripped-envelope CCSN has been identified and dubbed Type I superluminous supernovae (SLSNe). SLSNe can have luminosities up to 100 times brighter than SNe~Ib/c \citep{Quimby11}, and are therefore thought to be powered by a completely different mechanism, namely a millisecond magnetar engine created during the explosion \citep{Kasen10, Woosley10}.

Thanks to their extreme nature, SLSNe can be particularly useful for studying CCSNe, stellar evolution, and the high-redshift universe. Probably the most critical parameter to measure from SLSNe is their ejecta mass. Knowing the mass of the ejecta allows us to infer the masses of their progenitors, explore how these relate to other types of SNe, and test against different stellar evolution models. The ejecta masses of SLSNe can reach up to $\sim 40$ M$_\odot$ \citep{Blanchard20}, further than the maximum of $\sim 20$ M$_\odot$ inferred for typical SNe Ib/c \citep{Gomez22}. This suggests that the creation of a magnetar allows for the detection of SNe from more massive stars. Comparisons between the mass distribution of SLSNe to stellar evolution models seem to suggest that SLSNe are more likely to be produced in binary stars \citep{Blanchard20}. Other parameters such as the ejecta velocity or magnetar properties can help us to better understand the physics of the explosion. If we know the velocity of the ejecta, we can use this to estimate how efficient SLSNe are by measuring the total kinetic energy released in the explosion and comparing it to the total radiated luminosity. Additionally, measuring the magnetar spin period and magnetic field allows us to know the properties of the magnetar formed in the explosion, how much this contributes to the total luminosity of the SN, and whether or not magnetars can be formed in other types of SNe \citep{Gomez22}.

Thanks to their high luminosities, SLSNe can be detected out to larger distances than most SNe. Some of the most distant SLSNe have been found out to $z \sim 2$ \citep{Howell13, Hsu21}. These high-z SNe are likely the best probes we have to test different stellar evolution models and explore how the rotation rates, masses, and metallicities of massive stars evolve as a function of redshift (e.g., \citealt{Schulze18, Aguilera-Dena20}). Massive stars play a critical role in the reionization of the universe, and while individual stars are too faint to observe at $z \gtrsim 5$, SLSNe are luminous enough to be detected and allow for the direct characterization of massive star properties at these high redshifts during the epoch of reionization \citep{Moriya22}, and be used to estimate star formation rates as a function of redshift \citep{Frohmaier21}. Some studies have even suggested that SLSNe could be used as cosmological probes \citep{Scovacricchi16}. Although the validity of this technique is still pending, this could be tested with the discovery of even more distant SLSNe.

\section{SLSNe in the High Latitude Time Domain Survey}\label{sec:roman}

A significant component of the Nancy Grace Roman Space Telescope (\textit{Roman}) mission will be the High Latitude Time Domain Survey (HLTDS). While one of the key science drivers of the HLTDS is the study of Type Ia SNe, this survey will also provide us with an exquisite dataset for all types of transients. We advocate that the study of other types of extragalactic transients, such as SLSNe, should be considered when defining the parameters of the HLTDS. The HLTDS is expected to find hundreds of SLSNe during the lifetime of the mission \citep{Rose21, Moriya22}, significantly more than the $\sim 200$ total SLSNe discovered to date \citep{Gomez22, Chen23}. Thanks to the slowly-evolving nature of SLSNe, the HLTDS is particularly well suited to study them, without having to impose strong constraints on the cadence of the observations.

Here, we focus on optimizing the set of filters and cadences tht are best suited for the \textit{characterization} of SLSNe, to measure their physical parameters. We do not, however, consider the effects these choices have on their discovery rate or the ease of differentiating SLSNe from other transients. We consider the nominal HLTDS survey design from \cite{Rose21} as a starting point. This strategy proposes to observe the HLTDS fields on a 5 day cadence during the middle 2 years of the 5 year survey with two sets of filters: F062, F087, F106, and F129 in a wide region, and F106, F129, F158, and F184 in a deep overlapping region.

We estimate the signal-to-noise ratio (SNR) of the HLTDS observations using Pandeia 2.0 \citep{Pontoppidan16}, the \textit{Roman} exposure time calculator (ETC). We implement a strategy in which each field is observed with a 10s exposure time in a ``rapid" readout pattern, followed by a 300s exposure in a ``deep2" readout pattern with 9 groups. The addition of the ``rapid" readout pattern allows us to observe sources $\sim 2$ mag brighter without saturating the detectors. In this work, we implement this approximation to replicate the capability \textit{Roman} will have to observe with unevenly spaced resultants, which is not yet implemented in Pandeia. We assume a benchmark zodiacal background for these SNR measurements.

\section{Simulations}\label{sec:simulations}

\begin{figure}[h]
\begin{minipage}{0.5\textwidth}
  \includegraphics[width=\columnwidth]{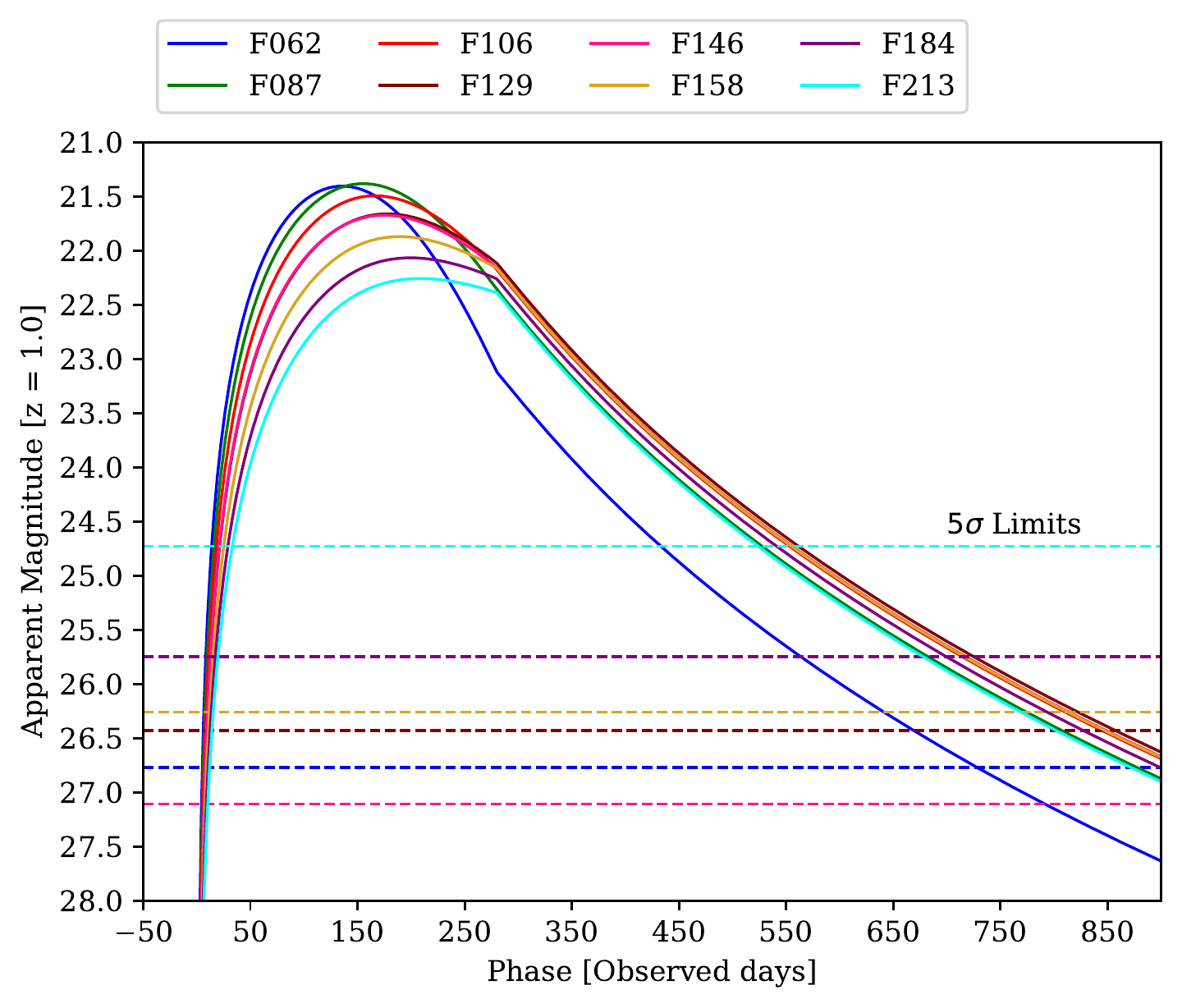}
\end{minipage}
\begin{minipage}{0.5\textwidth}
  \caption{Example light curve model of a SLSN generated with {\tt MOSFiT} at a redshift of $z = 1.0$, where phase is observed days from explosion. The nominal $5\sigma$ upper limits of the HLTDS for a 300s in each \textit{Roman} filter are shown as dashed horizontal lines. \label{fig:model}}
\end{minipage}
\end{figure}

For this analysis, we simulate SLSNe light curves powered by a magnetar central engine model created with the Modular Open-Source Fitter for Transients ({\tt MOSFiT}) Python package, a flexible code that uses a Markov chain Monte Carlo (MCMC) implementation to fit the multi-band light curves of transients \citep{guillochon18}. We adopt the mean physical parameters of SLSNe from \cite{Nicholl17} and create a model with an ejecta mass of $M_{\rm ej} = 9.0$ M$_\odot$, an ejecta velocity of $V_{\rm ej} = 5000$ km s$^{-1}$, a magnetar spin period of $P_s = 2.0$ ms, and a magnetic field of $B_\perp = 0.5 \times 10^{14}$G. We then calculate the expected magnitude of the SLSNe in the \textit{Roman} filters at redshifts of $z = 0.1$, 1.0, 3.0, and 5.0. An example of a model light curve is shown in Figure~\ref{fig:model}. 

\begin{figure}
    \begin{center}
        \includegraphics[width=0.4\columnwidth]{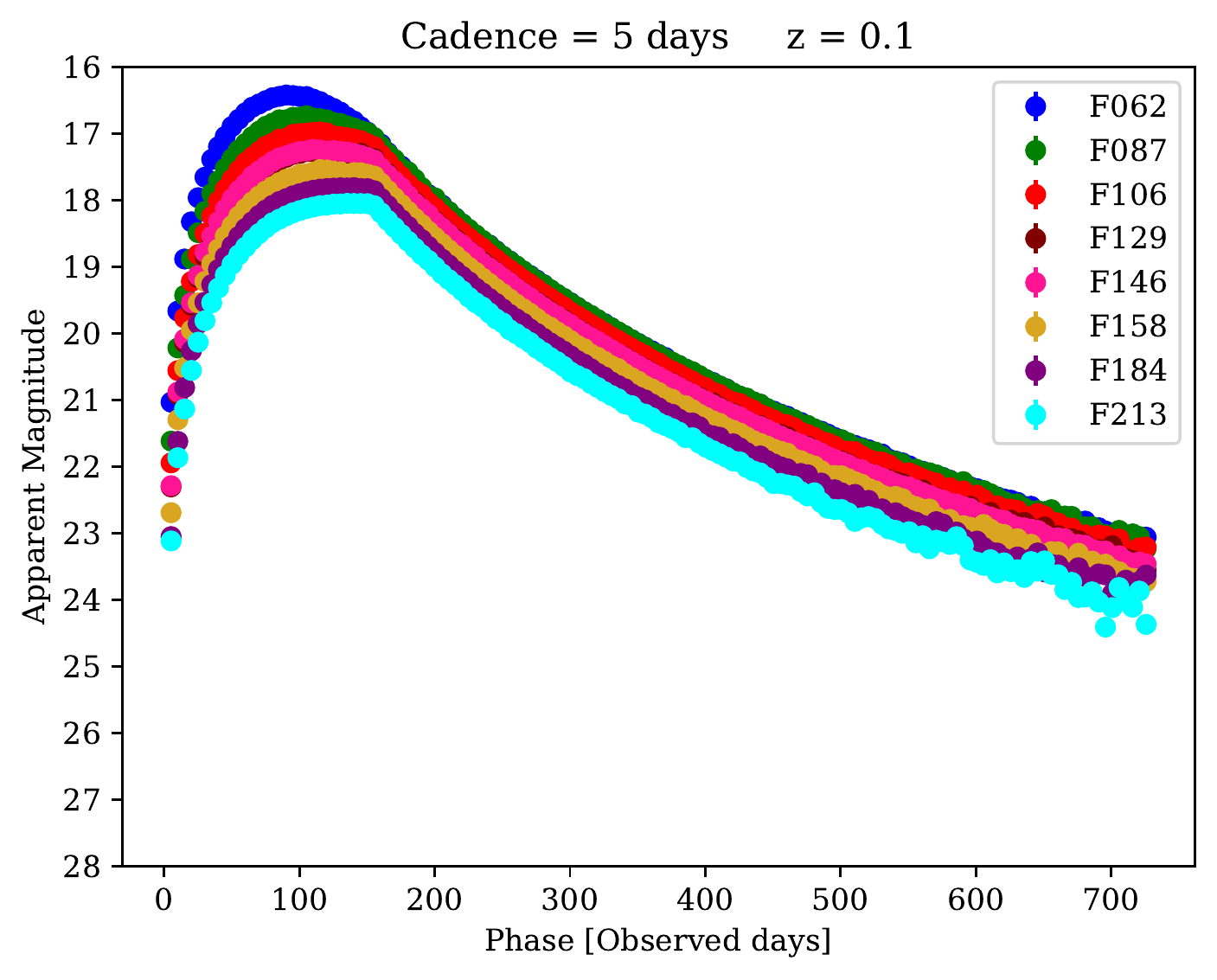}
        \includegraphics[width=0.4\columnwidth]{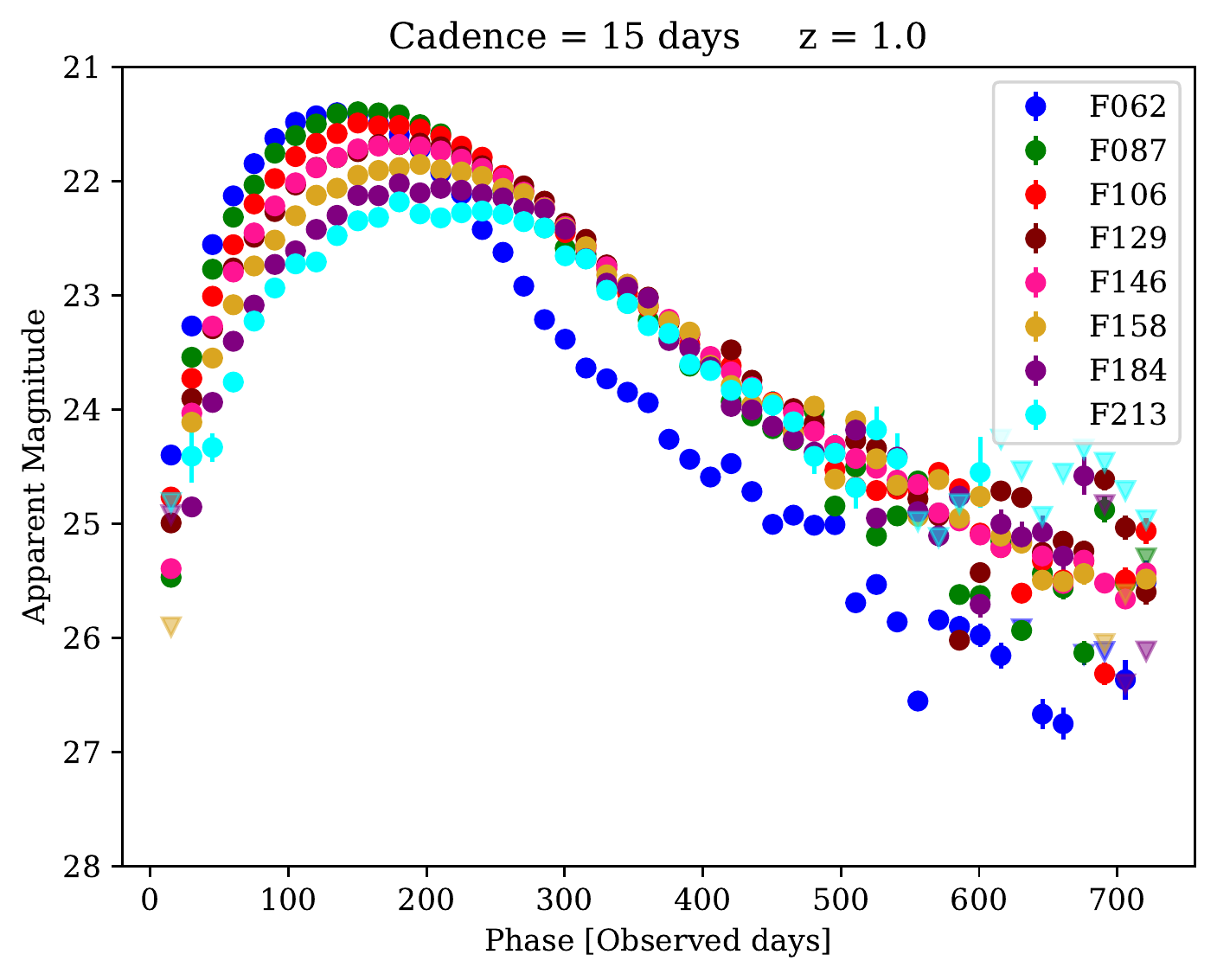}
        \includegraphics[width=0.4\columnwidth]{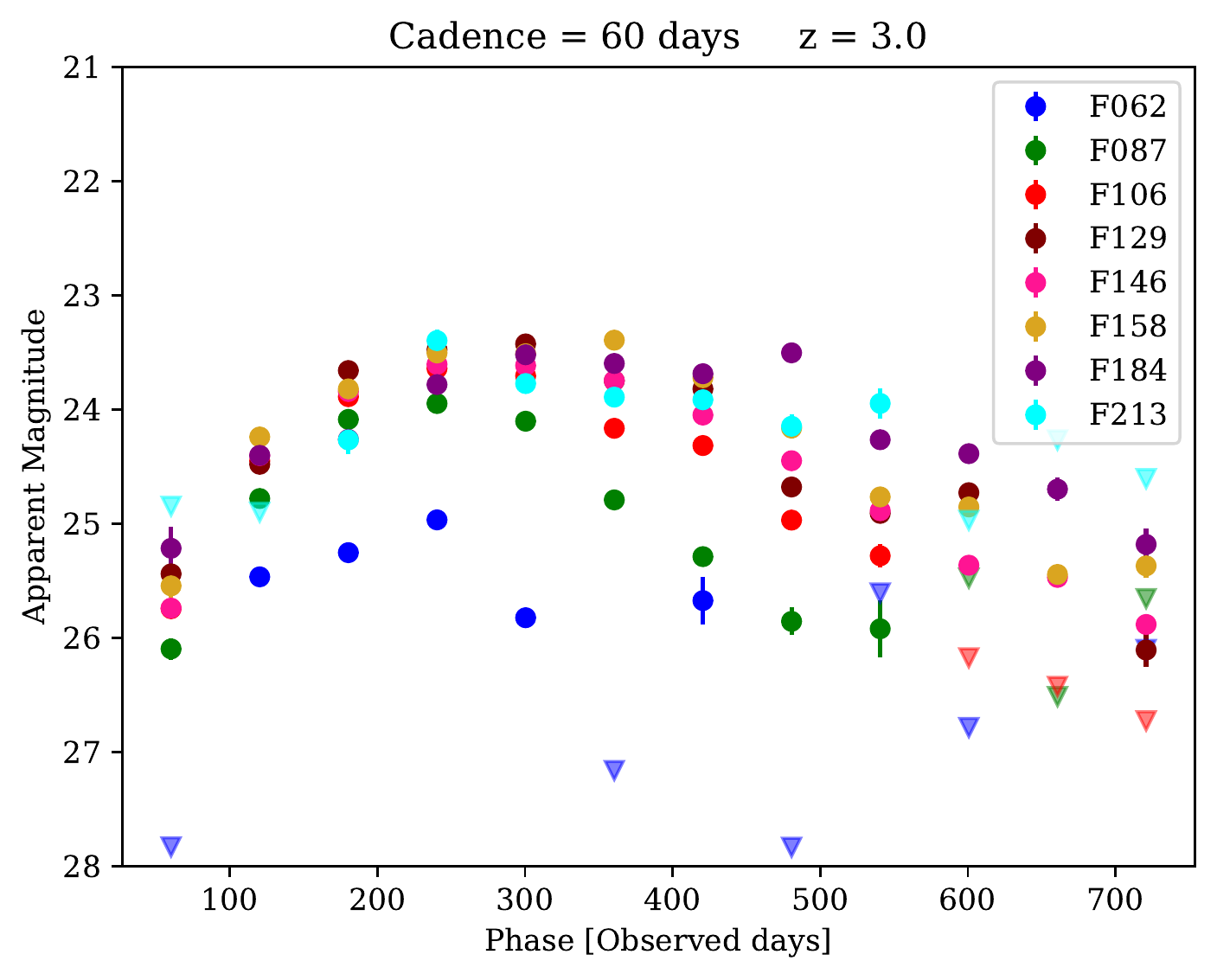}
        \includegraphics[width=0.4\columnwidth]{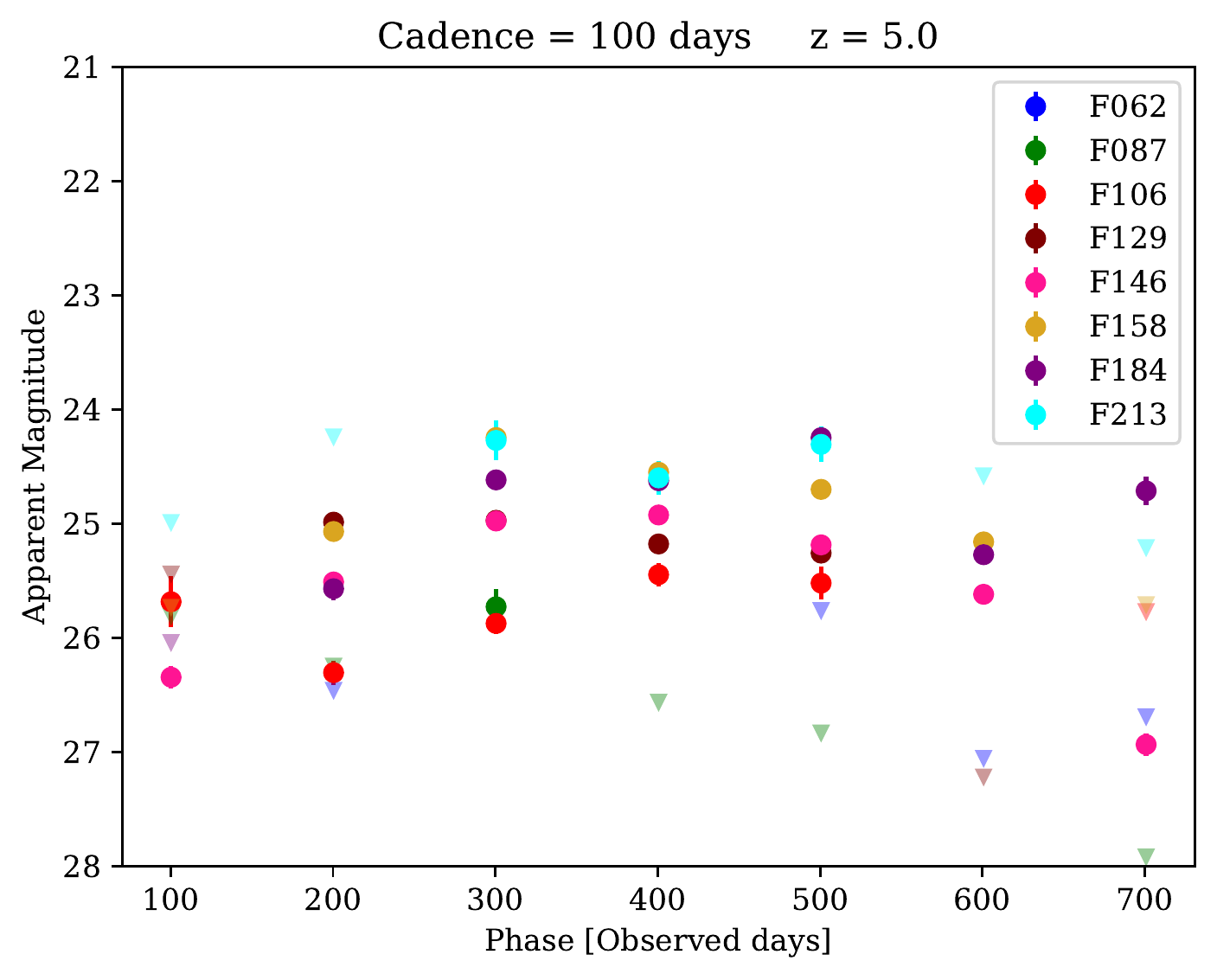}
        \caption{Representative examples of four simulated HLTDS observations of SLSN models at increasingly higher redshifts and longer cadences, where phase is observed days from explosion. At the highest redshifts the SNe become dimmer and harder to detect, but also last longer, delaying the time of peak. Triangles denote $5 \sigma$ non-detections. \label{fig:lcs}}
    \end{center}
\end{figure}

We add noise to the model light curves based on the SNR estimates derived using Pandeia, and convert any measurements below the $5\sigma$ detection threshold to upper limits. We simulate light curves at cadences of 5, 15, 30, 45, 60, 80, and 100 days, assuming the best-case scenario that the SN exploded at the beginning of the survey. We simulate the light curves using six different filter combinations, listed in Table~\ref{tab:filters} and some representative light curve examples shown in Figure~\ref{fig:lcs}. We include the set of ``Nominal 6" filters defined in \cite{Rose21}, the ``Bluest 4" set has the four blue filters from the wide component of the survey design, and ``Reddest 4" has the four red filters from the deep component of the survey. We also include a model with all \textit{Roman} filters, and one with just the bluest and reddest filters (F062 + F213).

\begin{deluxetable*}{cc}
    \tablecaption{Names for sets of filters explored \label{tab:filters}}
    \tablewidth{0pt}
    \tablehead{
        \colhead{Name} & \colhead{Filters}}
    \startdata
    All              & F062+F087+F106+F129+F146+F158+F184+F213 \\
    Nominal 6 + F213 & F062+F087+F106+F129+F158+F184+F213      \\
    Nominal 6        & F062+F087+F106+F129+F158+F184           \\
    Bluest 4         & F062+F087+F106+F129                     \\
    Reddest 4        & F106+F129+F158+F184                     \\
    F062 + F213      & F062+F213                               \\
    \enddata
    \tablecomments{The set of ``Nominal 6" filters are the ones defined in \cite{Rose21}, where ``Bluest 4" are the four blue filters from the wide component of the survey and ``Reddest 4" are the four red filters from the deep component of the survey.}
\end{deluxetable*}

\section{Results}\label{sec:results}

To determine how accurately we can recover the physical parameters of SLSNe under different observing strategies, we fit the simulated model light curves using the same magnetar central engine model with {\tt MOSFiT}. We assume that the redshift of the SN is known, and that there is no intrinsic host extinction; which is usually the case for the dim dwarf galaxies that host SLSNe \citep{Lunnan15}. To maintain a realistic scenario, we do not assume that any other parameter in the {\tt MOSFiT} model is known. The parameters we fit for include: ejecta mass $M_{\text{ej}}$, ejecta velocity $V_{\text{ej}}$, neutron star mass $M_{\text{NS}}$, magnetar spin period $P_{s}$, magnetar magnetic field strength $B_{\perp}$, angle of the dipole moment $\theta_{\text{BP}}$, explosion time relative to first data point $t_{\text{exp}}$, photosphere temperature floor $T_{\text{min}}$, optical opacity $\kappa$, and gamma-ray opacity $\kappa_{\gamma}$.

\begin{figure}
    \begin{center}
        \includegraphics[width=\columnwidth]{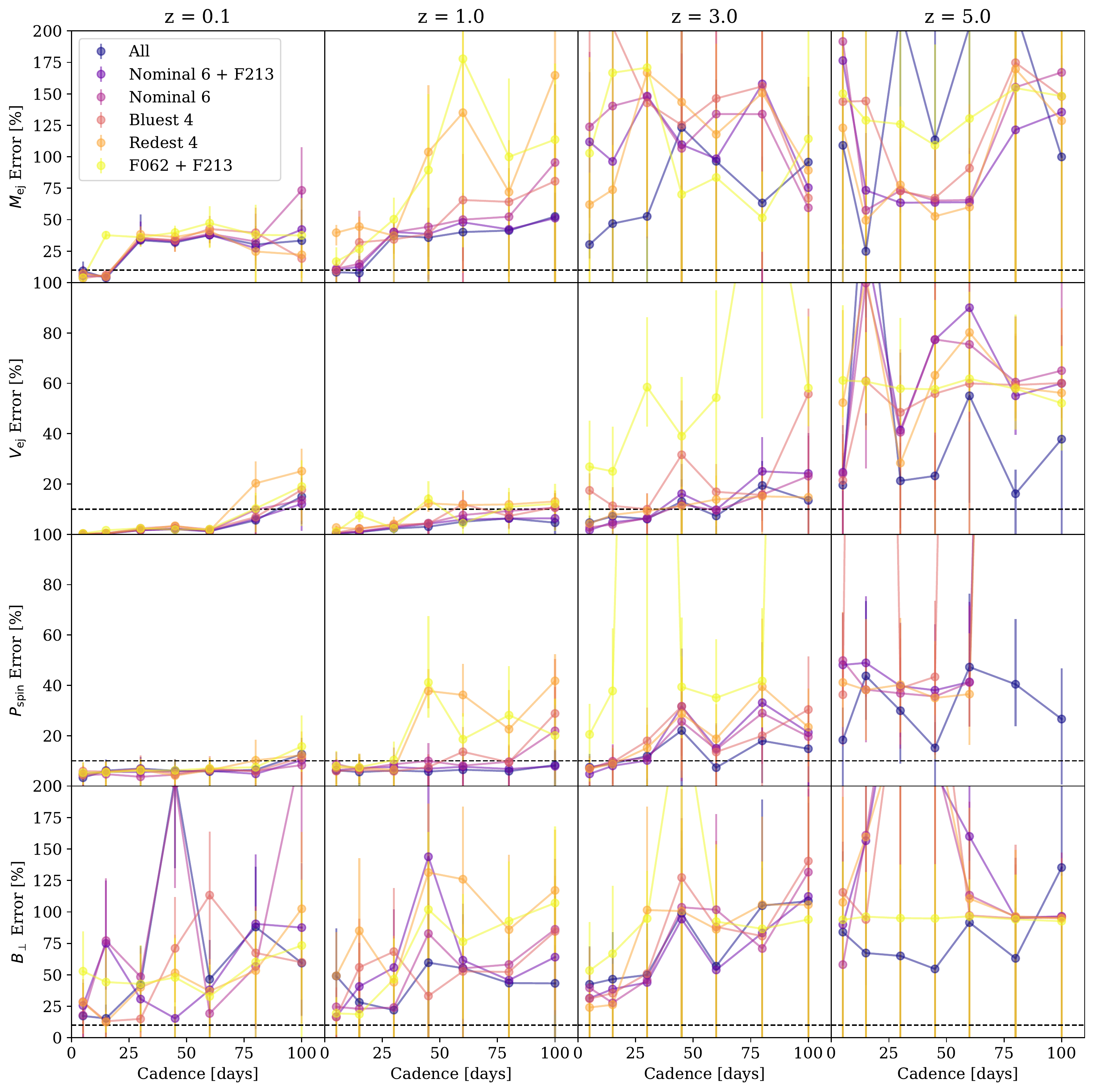}
        \caption{Error in the measurement of the ejecta mass, ejecta velocity, spin period, and magnetic field as a function of survey cadence for four different redshifts (z = 0.1, 1.0, 3.0, 5.0). The colored lines represent different filter choices, and the error bars represent the scatter on 150 individual model realizations. \label{fig:parameters}}
    \end{center}
\end{figure}

The most critical parameter we aim to recover is the ejecta mass $M_{\text{ej}}$, which gives us some of the most critical information about the progenitor that produced the SN. In Figure~\ref{fig:parameters}, we show how the uncertainty on the measurement of $M_{\text{ej}}$ depends on filter choice and cadence at various redshifts. We find that for the most nearby SLSNe, a cadence of $\lesssim 30$ days is needed to measure $M_{\text{ej}}$ with an uncertainty of less than 10\%. With the exception of the under-performing F062 + F213 filter set, all other filter choices provide similar constraints on the value of $M_{\text{ej}}$.

We find that the cadence choice is extremely important to determine $V_{\text{ej}}$. In the second row of Figure~\ref{fig:parameters} we show how the uncertainty in the measurement of $V_{\text{ej}}$ drastically increases for cadences $\gtrsim 70$ days for the SLSNe at $z \sim 0.1$. We are otherwise able to measure $V_{\text{ej}}$ with a 10\% uncertainty for most filter choices and cadences up to $z = 3$, with the exception of the F062 + F213 filter set, which produces much worse estimates above $z = 3$.

For both $M_{\text{ej}}$ and $V_{\text{ej}}$ we find that observing with all \textit{Roman} filters reduces the measurement in the uncertainty by a factor of $\sim 2$ compared to the ``Nominal 6" filters for SLSNe between $z = 1.0$ and $z = 3.0$, regardless of the cadence of the observations. For $P_{s}$ we find that the Reddest 4 and F062+F213 filter choices perform worse than others for SLSNe at $z = 1$. With the exception of some fast cadences at $z = 0.1$, we find that most of these strategies are unable to constrain $B_{\perp}$ with less than 10\% error.

\subsection{High Redshift SLSNe}\label{sec:high-z}

We have determined from our analysis the set of filters and cadences that allow us to measure the physical parameters of SLSNe with a 10\% accuracy. Nevertheless, we find that for SLSNe at $z = 5$ (critical for the study of the epoch of reionization and the redshift-dependence of stellar evolution), we are not able to recover the values of $M_{\text{ej}}$ or $V_{\text{ej}}$ to better than 10\% with any choice of filter set or cadence. Therefore, in this section, we relax some of the assumptions from the previous section. We test the possibility of extending the duration of the survey, increasing the exposure time of the observations, or considering only the most luminous SLSNe. For these tests, we only consider cadences of 30 days or larger given that at this redshift SLSNe will be heavily time-dilated and thus do not require rapid cadences. We also only consider the three sets of filter choices from \cite{Rose21}.

First, we explore the effects of doubling the exposure time of the ``deep2" component described in \S\ref{sec:roman} to a total of 600s per filter. This strategy reduces the uncertainty in the measurement of $V_{\rm ej}$ and $P_s$ at all cadences by a factor of $\sim 3$ and $\sim 5$, respectively. The uncertainty on the measurement of $B_\perp$ is also reduced by a similar factor, but not below the target 10\% error. The uncertainty on the measurement of $M_{\rm ej}$ is not significantly different.

We perform the same simulations as above, but instead of adopting the mean parameters of SLSNe, we adopt those of iPTF13ajg, one of the most luminous SLSNe ever discovered \cite{Vreeswijk14}. We create SLSN models using an ejecta mass of $M_{\rm ej} = 40.0$ M$_\odot$, an ejecta velocity of $V_{\rm ej} = 10000$ km s$^{-1}$, a magnetar spin period of $P_s = 2.0$ ms, and a magnetic field of $B_\perp = 1.3 \times 10^{14}$G. Any cadence or filter choice results in an uncertainty in $M_{\rm ej}$ of $\sim 75$\%, the uncertainty in the other three parameters are reduced by a factor of $2 - 5$.

Finally, we explore the effects of doubling the length of the HLTDS to four years instead of two. This is the strategy that results in the most accurate estimates for $P_s$ and $V_{\rm ej}$ for effectively all cadences. The uncertainty in $B_\perp$ is significantly reduced down to $\sim 30$\% for cadences below 60 days. Similarly to doubling the exposure time, the estimate on $M_{\rm ej}$ does not appear to be significantly affected. Doubling the exposure time for all filters would also double the required time to complete the HLTDS. On the other hand, doubling the duration of the HLTDS at a modest cadence of 60 days represents only an $\sim 8$\% increase over the nominal survey design. Given that the effects of doubling the exposure times or doubling the length of the HLTDS are comparable, the latter is much preferred.

\begin{figure}
    \begin{center}
        \includegraphics[width=\columnwidth]{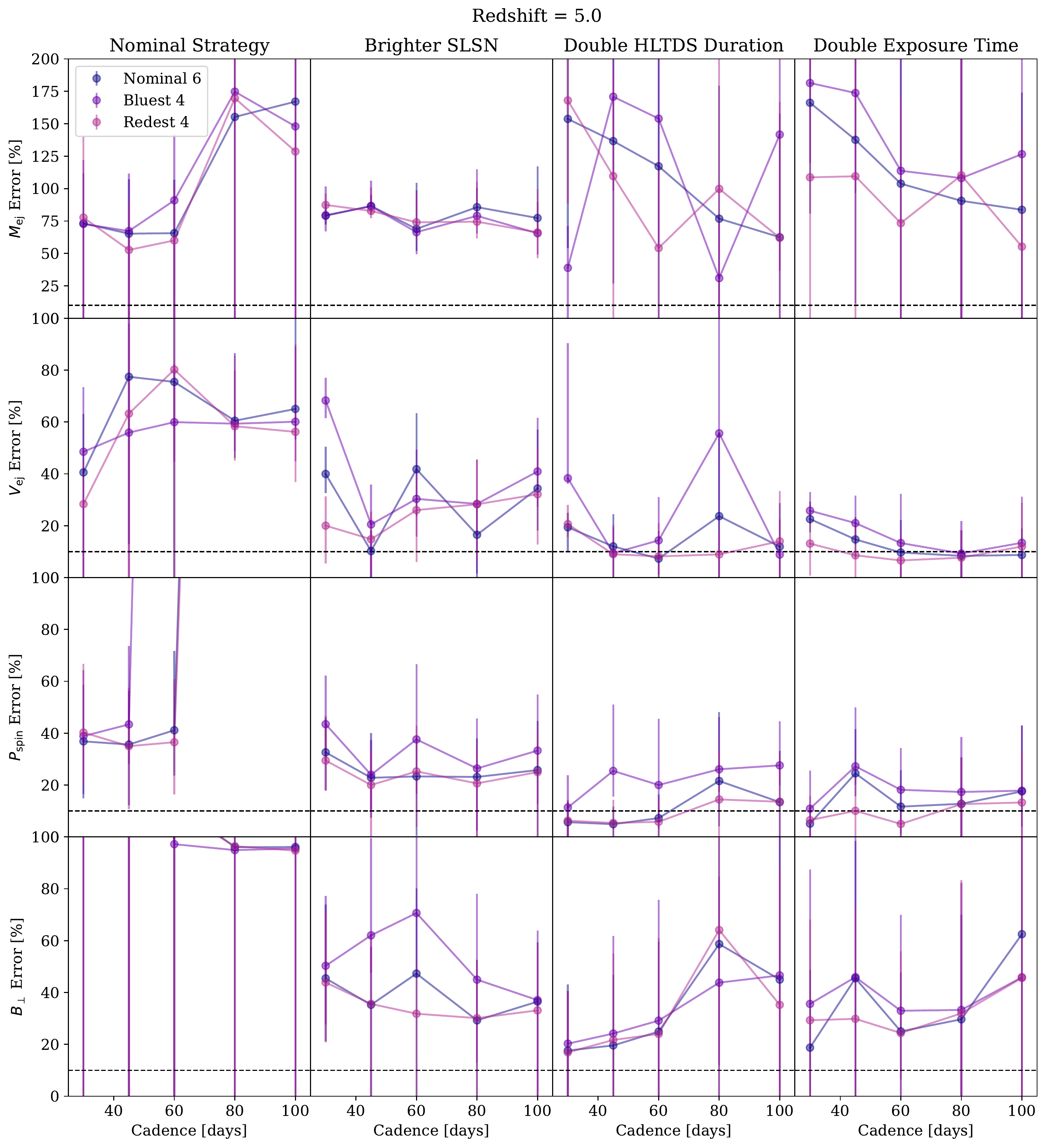}
        \caption{Error in the measurement of the ejecta mass, ejecta velocity, spin period, and magnetic field as a function of survey cadence for different HLTDS strategies. The Nominal Strategy is the same as the one shown in Figure~\ref{fig:parameters}. The Brighter SLSN strategy uses the physical parameters of iPTF13ajg as opposed to the mean SLSNe parameters. The Double HLTDS Duration strategy doubles the duration to a total of 4 years. The Double Exposure Time doubles the exposure time of each image to 600s. The colored lines represent different filter choices, and the error bars represent the scatter on 150 individual model realizations \label{fig:strategies}}
    \end{center}
\end{figure}

\section{Conclusion}\label{sec:conclusion}

We have simulated observations of SLSNe in the HLTDS to determine the set of filters and cadences that will allow us to characterize their physical parameters with a target uncertainty of 10\% and found:
\begin{itemize}
    \item Four filters are sufficient to accurately characterize SLSNe out to $z = 1$.
    \item Six filters are preferred to characterize SLSNe out to $z = 3$.
    \item A cadence of $< 20$ days is required to constrain $M_{\rm ej}$, $V_{\rm ej}$, and $P_s$ with better than 10\% uncertainty.
    \item For cadences above 70 days, the uncertainty in all parameters, but mainly $V_{\rm ej}$, increases by a factor of 2 to 5.
    \item Doubling the duration of the HLTDS, even with a 60 day cadence, can reduce the uncertainty in $V_{\rm ej}$, $P_s$, and $B_\perp$ enough to go below 10\% for SLSNe at $z = 5$.
    \item If the duration of the HLTDS is doubled, the Reddest 4 filters perform better than the Bluest 4.
    \item Doubling the exposure time of the survey can improve the estimates of $V_{\rm ej}$, $P_s$, and $B_\perp$ for SLSNe at $z = 5$ by a factor of $\sim 5$; although at a high cost of exposure time.
    \item For SLSNe at $z = 5$, the nominal survey strategy will only be able to characterize the most luminous SNe.
\end{itemize}

\bibliography{references}

\begin{thebibliography}{}
\expandafter\ifx\csname natexlab\endcsname\relax\def\natexlab#1{#1}\fi
\providecommand{\url}[1]{\href{#1}{#1}}
\providecommand{\dodoi}[1]{doi:~\href{http://doi.org/#1}{\nolinkurl{#1}}}

\bibitem[{{Aguilera-Dena} {et~al.}(2020){Aguilera-Dena}, {Langer},
  {Antoniadis}, \& {M{\"u}ller}}]{Aguilera-Dena20}
{Aguilera-Dena}, D.~R., {Langer}, N., {Antoniadis}, J., \& {M{\"u}ller}, B.
  2020,
  \hypersetup{urlcolor=magenta}\href{https://dx.doi.org/10.3847/1538-4357/abb138}{Astrophysical
  Journal},
  \hypersetup{urlcolor=blue}\href{https://ui.adsabs.harvard.edu/abs/2020ApJ...901..114A}{901,
  114}

\bibitem[{{Aguilera-Dena} {et~al.}(2023){Aguilera-Dena}, {M{\"u}ller},
  {Antoniadis}, {Langer}, {Dessart}, {Vigna-G{\'o}mez}, \&
  {Yoon}}]{Aguilera-Dena23}
{Aguilera-Dena}, D.~R., {M{\"u}ller}, B., {Antoniadis}, J., {et~al.} 2023,
  \hypersetup{urlcolor=magenta}\href{https://dx.doi.org/10.1051/0004-6361/202243519}{Astronomy
  \& Astrophysics},
  \hypersetup{urlcolor=blue}\href{https://ui.adsabs.harvard.edu/abs/2023A&A...671A.134A}{671,
  A134}

\bibitem[{{Blanchard} {et~al.}(2020){Blanchard}, {Berger}, {Nicholl}, \&
  {Villar}}]{Blanchard20}
{Blanchard}, P.~K., {Berger}, E., {Nicholl}, M., \& {Villar}, V.~A. 2020,
  \hypersetup{urlcolor=magenta}\href{https://dx.doi.org/10.3847/1538-4357/ab9638}{\apj},
  \hypersetup{urlcolor=blue}\href{https://ui.adsabs.harvard.edu/abs/2020ApJ...897..114B}{897,
  114}

\bibitem[{{Chen} {et~al.}(2023){Chen}, {Yan}, {Kangas}, {Lunnan}, {Schulze},
  {Sollerman}, {Perley}, {Chen}, {Taggart}, {Hinds}, {Gal-Yam}, {Wang},
  {Andreoni}, {Bellm}, {Bloom}, {Burdge}, {Burgos}, {Cook}, {Dahiwale}, {De},
  {Dekany}, {Dugas}, {Frederik}, {Fremling}, {Graham}, {Hankins}, {Ho},
  {Jencson}, {Karambelkar}, {Kasliwal}, {Kulkarni}, {Laher}, {Rusholme},
  {Sharma}, {Taddia}, {Tartaglia}, {Thomas}, {Tzanidakis}, {Van Roestel},
  {Walter}, {Yang}, {Yao}, \& {Yaron}}]{Chen23}
{Chen}, Z.~H., {Yan}, L., {Kangas}, T., {et~al.} 2023,
  \hypersetup{urlcolor=magenta}\href{https://dx.doi.org/10.3847/1538-4357/aca161}{\apj},
  \hypersetup{urlcolor=blue}\href{https://ui.adsabs.harvard.edu/abs/2023ApJ...943...41C}{943,
  41}

\bibitem[{{Filippenko}(1997)}]{Filippenko97}
{Filippenko}, A.~V. 1997,
  \hypersetup{urlcolor=magenta}\href{https://dx.doi.org/10.1146/annurev.astro.35.1.309}{\araa},
  \hypersetup{urlcolor=blue}\href{https://ui.adsabs.harvard.edu/abs/1997ARA&A..35..309F}{35,
  309}

\bibitem[{{Frohmaier} {et~al.}(2021){Frohmaier}, {Angus}, {Vincenzi},
  {Sullivan}, {Smith}, {Nugent}, {Cenko}, {Gal-Yam}, {Kulkarni}, {Law}, \&
  {Quimby}}]{Frohmaier21}
{Frohmaier}, C., {Angus}, C.~R., {Vincenzi}, M., {et~al.} 2021,
  \hypersetup{urlcolor=magenta}\href{https://dx.doi.org/10.1093/mnras/staa3607}{\mnras},
  \hypersetup{urlcolor=blue}\href{https://ui.adsabs.harvard.edu/abs/2021MNRAS.500.5142F}{500,
  5142}

\bibitem[{{Gomez} {et~al.}(2022){Gomez}, {Berger}, {Nicholl}, {Blanchard}, \&
  {Hosseinzadeh}}]{Gomez22}
{Gomez}, S., {Berger}, E., {Nicholl}, M., {Blanchard}, P.~K., \&
  {Hosseinzadeh}, G. 2022,
  \hypersetup{urlcolor=magenta}\href{https://dx.doi.org/10.3847/1538-4357/ac9842}{\apj},
  \hypersetup{urlcolor=blue}\href{https://ui.adsabs.harvard.edu/abs/2022ApJ...941..107G}{941,
  107}

\bibitem[{{Guillochon} {et~al.}(2018){Guillochon}, {Nicholl}, {Villar},
  {Mockler}, {Narayan}, {Mandel}, {Berger}, \& {Williams}}]{guillochon18}
{Guillochon}, J., {Nicholl}, M., {Villar}, V.~A., {et~al.} 2018,
  \hypersetup{urlcolor=magenta}\href{https://dx.doi.org/10.3847/1538-4365/aab761}{\apjs},
  \hypersetup{urlcolor=blue}\href{https://ui.adsabs.harvard.edu/abs/2018ApJS..236....6G}{236,
  6}

\bibitem[{{Howell} {et~al.}(2013){Howell}, {Kasen}, {Lidman}, {Sullivan},
  {Conley}, {Astier}, {Balland}, {Carlberg}, {Fouchez}, {Guy}, {Hardin},
  {Pain}, {Palanque-Delabrouille}, {Perrett}, {Pritchet}, {Regnault}, {Rich},
  \& {Ruhlmann-Kleider}}]{Howell13}
{Howell}, D.~A., {Kasen}, D., {Lidman}, C., {et~al.} 2013,
  \hypersetup{urlcolor=magenta}\href{https://dx.doi.org/10.1088/0004-637X/779/2/98}{\apj},
  \hypersetup{urlcolor=blue}\href{https://ui.adsabs.harvard.edu/abs/2013ApJ...779...98H}{779,
  98}

\bibitem[{{Hsu} {et~al.}(2021){Hsu}, {Hosseinzadeh}, \& {Berger}}]{Hsu21}
{Hsu}, B., {Hosseinzadeh}, G., \& {Berger}, E. 2021,
  \hypersetup{urlcolor=magenta}\href{https://dx.doi.org/10.3847/1538-4357/ac1aca}{\apj},
  \hypersetup{urlcolor=blue}\href{https://ui.adsabs.harvard.edu/abs/2021ApJ...921..180H}{921,
  180}

\bibitem[{{Kasen} \& {Bildsten}(2010)}]{Kasen10}
{Kasen}, D., \& {Bildsten}, L. 2010,
  \hypersetup{urlcolor=magenta}\href{https://dx.doi.org/10.1088/0004-637X/717/1/245}{\apj},
  \hypersetup{urlcolor=blue}\href{https://ui.adsabs.harvard.edu/abs/2010ApJ...717..245K}{717,
  245}

\bibitem[{{Lunnan} {et~al.}(2015){Lunnan}, {Chornock}, {Berger}, {Rest},
  {Fong}, {Scolnic}, {Jones}, {Soderberg}, {Challis}, \& {Drout}}]{Lunnan15}
{Lunnan}, R., {Chornock}, R., {Berger}, E., {et~al.} 2015,
  \hypersetup{urlcolor=magenta}\href{https://dx.doi.org/10.1088/0004-637X/804/2/90}{\apj},
  \hypersetup{urlcolor=blue}\href{https://ui.adsabs.harvard.edu/abs/2015ApJ...804...90L}{804,
  90}

\bibitem[{{Moriya} {et~al.}(2022){Moriya}, {Quimby}, \& {Robertson}}]{Moriya22}
{Moriya}, T.~J., {Quimby}, R.~M., \& {Robertson}, B.~E. 2022,
  \hypersetup{urlcolor=magenta}\href{https://dx.doi.org/10.3847/1538-4357/ac415e}{\apj},
  \hypersetup{urlcolor=blue}\href{https://ui.adsabs.harvard.edu/abs/2022ApJ...925..211M}{925,
  211}

\bibitem[{{Nicholl} {et~al.}(2017){Nicholl}, {Guillochon}, \&
  {Berger}}]{Nicholl17}
{Nicholl}, M., {Guillochon}, J., \& {Berger}, E. 2017,
  \hypersetup{urlcolor=magenta}\href{https://dx.doi.org/10.3847/1538-4357/aa9334}{\apj},
  \hypersetup{urlcolor=blue}\href{https://ui.adsabs.harvard.edu/abs/2017ApJ...850...55N}{850,
  55}

\bibitem[{{Podsiadlowski} {et~al.}(1992){Podsiadlowski}, {Joss}, \&
  {Hsu}}]{Podsiadlowski92}
{Podsiadlowski}, P., {Joss}, P.~C., \& {Hsu}, J.~J.~L. 1992,
  \hypersetup{urlcolor=magenta}\href{https://dx.doi.org/10.1086/171341}{Astrophysical
  Journal},
  \hypersetup{urlcolor=blue}\href{https://ui.adsabs.harvard.edu/abs/1992ApJ...391..246P}{391,
  246}

\bibitem[{{Pontoppidan} {et~al.}(2016){Pontoppidan}, {Pickering}, {Laidler},
  {Gilbert}, {Sontag}, {Slocum}, {Sienkiewicz}, {Hanley}, {Earl}, {Pueyo},
  {Ravindranath}, {Karakla}, {Robberto}, {Noriega-Crespo}, \&
  {Barker}}]{Pontoppidan16}
{Pontoppidan}, K.~M., {Pickering}, T.~E., {Laidler}, V.~G., {et~al.} 2016, in
  Society of Photo-Optical Instrumentation Engineers (SPIE) Conference Series,
  Vol. 9910, Observatory Operations: Strategies, Processes, and Systems VI, ed.
  A.~B. {Peck}, R.~L. {Seaman}, \& C.~R. {Benn}, 991016

\bibitem[{{Quimby} {et~al.}(2011){Quimby}, {Kulkarni}, {Kasliwal}, {Gal-Yam},
  {Arcavi}, {Sullivan}, {Nugent}, {Thomas}, {Howell}, \& {Nakar}}]{Quimby11}
{Quimby}, R.~M., {Kulkarni}, S.~R., {Kasliwal}, M.~M., {et~al.} 2011,
  \hypersetup{urlcolor=magenta}\href{https://dx.doi.org/10.1038/nature10095}{\nat},
  \hypersetup{urlcolor=blue}\href{https://ui.adsabs.harvard.edu/abs/2011Natur.474..487Q}{474,
  487}

\bibitem[{{Rose} {et~al.}(2021){Rose}, {Baltay}, {Hounsell}, {Macias}, {Rubin},
  {Scolnic}, {Aldering}, {Bohlin}, {Dai}, {Deustua}, {Foley}, {Fruchter},
  {Galbany}, {Jha}, {Jones}, {Joshi}, {Kelly}, {Kessler}, {Kirshner}, {Mandel},
  {Perlmutter}, {Pierel}, {Qu}, {Rabinowitz}, {Rest}, {Riess}, {Rodney},
  {Sako}, {Siebert}, {Strolger}, {Suzuki}, {Thorp}, {Van Dyk}, {Wang}, {Ward},
  \& {Wood-Vasey}}]{Rose21}
{Rose}, B.~M., {Baltay}, C., {Hounsell}, R., {et~al.} 2021,
  \hypersetup{urlcolor=magenta}\href{https://dx.doi.org/10.48550/arXiv.2111.03081}{arXiv
  e-prints},
  \hypersetup{urlcolor=magenta}\href{https://arxiv.org/abs/2111.03081}{arXiv}{:}\hypersetup{urlcolor=blue}\href{https://ui.adsabs.harvard.edu/abs/2021arXiv211103081R}{2111.03081}

\bibitem[{{Schulze} {et~al.}(2018){Schulze}, {Kr{\"u}hler}, {Leloudas},
  {Gorosabel}, {Mehner}, {Buchner}, {Kim}, {Ibar}, {Amor{\'\i}n},
  {Herrero-Illana}, {Anderson}, {Bauer}, {Christensen}, {de Pasquale}, {de
  Ugarte Postigo}, {Gallazzi}, {Hjorth}, {Morrell}, {Malesani}, {Sparre},
  {Stalder}, {Stark}, {Th{\"o}ne}, \& {Wheeler}}]{Schulze18}
{Schulze}, S., {Kr{\"u}hler}, T., {Leloudas}, G., {et~al.} 2018,
  \hypersetup{urlcolor=magenta}\href{https://dx.doi.org/10.1093/mnras/stx2352}{\mnras},
  \hypersetup{urlcolor=blue}\href{https://ui.adsabs.harvard.edu/abs/2018MNRAS.473.1258S}{473,
  1258}

\bibitem[{{Scovacricchi} {et~al.}(2016){Scovacricchi}, {Nichol}, {Bacon},
  {Sullivan}, \& {Prajs}}]{Scovacricchi16}
{Scovacricchi}, D., {Nichol}, R.~C., {Bacon}, D., {Sullivan}, M., \& {Prajs},
  S. 2016,
  \hypersetup{urlcolor=magenta}\href{https://dx.doi.org/10.1093/mnras/stv2752}{\mnras},
  \hypersetup{urlcolor=blue}\href{https://ui.adsabs.harvard.edu/abs/2016MNRAS.456.1700S}{456,
  1700}

\bibitem[{{Vreeswijk} {et~al.}(2014){Vreeswijk}, {Savaglio}, {Gal-Yam}, {De
  Cia}, {Quimby}, {Sullivan}, {Cenko}, {Perley}, {Filippenko}, {Clubb},
  {Taddia}, {Sollerman}, {Leloudas}, {Arcavi}, {Rubin}, {Kasliwal}, {Cao},
  {Yaron}, {Tal}, {Ofek}, {Capone}, {Kutyrev}, {Toy}, {Nugent}, {Laher},
  {Surace}, \& {Kulkarni}}]{Vreeswijk14}
{Vreeswijk}, P.~M., {Savaglio}, S., {Gal-Yam}, A., {et~al.} 2014,
  \hypersetup{urlcolor=magenta}\href{https://dx.doi.org/10.1088/0004-637X/797/1/24}{\apj},
  \hypersetup{urlcolor=blue}\href{https://ui.adsabs.harvard.edu/abs/2014ApJ...797...24V}{797,
  24}

\bibitem[{{Woosley}(2010)}]{Woosley10}
{Woosley}, S.~E. 2010,
  \hypersetup{urlcolor=magenta}\href{https://dx.doi.org/10.1088/2041-8205/719/2/L204}{\apj},
  \hypersetup{urlcolor=blue}\href{https://ui.adsabs.harvard.edu/abs/2010ApJ...719L.204W}{719,
  L204}

\bibitem[{{Woosley} {et~al.}(2007){Woosley}, {Blinnikov}, \&
  {Heger}}]{Woosley07}
{Woosley}, S.~E., {Blinnikov}, S., \& {Heger}, A. 2007,
  \hypersetup{urlcolor=magenta}\href{https://dx.doi.org/10.1038/nature06333}{\nat},
  \hypersetup{urlcolor=blue}\href{https://ui.adsabs.harvard.edu/abs/2007Natur.450..390W}{450,
  390}

\end{thebibliography}

\end{document}